\documentclass[manuscript]{aa}

\usepackage{natbib}
\usepackage[varg]{txfonts}
\usepackage{graphicx}

\bibpunct{(}{)}{;}{a}{}{,} % to follow the A&A style

\begin{document}

\title{Photophoresis boosts giant planet formation}

\author{Jens Teiser\inst{1} \and Sarah E. Dodson-Robinson \inst{2}}

\institute{Faculty of Physics, University of Duisburg-Essen, Lotharstr. 1, D-47048 Duisburg, Germany \and The University of Texas at Austin, Department of Astronomy, 2515 Speedway Dr. Stop 1400, Austin, TX 78712, USA}

\date{}

%\email{jens.teiser@uni-due.de}
%\and 

%\author{S. Dodson-Robinson}

%\affil{The University of Texas at Austin, Department of
%Astronomy, 2515 Speedway Dr. Stop 1400, Austin, TX 78712, USA}
%\email{sdr@astro.as.utexas.edu}

\abstract{In the core accretion model of giant planet formation, a solid
protoplanetary core begins to accrete gas directly from the nebula when
its mass reaches $\sim 5 M_{\oplus}$. The protoplanet has at most a few
million years to reach runaway gas accretion, as young stars lose their
gas disks after 10 million years at the latest.  Yet gas accretion also
brings small dust grains entrained in the gas into the planetary
atmosphere. Dust accretion creates an optically thick protoplanetary atmosphere that
cannot efficiently radiate away the kinetic energy deposited by incoming
planetesimals. A dust-rich atmosphere severely slows down atmospheric
cooling, contraction, and inflow of new gas, in contradiction to the
observed timescales of planet formation.  Here we show that
photophoresis is a strong mechanism for pushing dust out of the
planetary atmosphere due to the momentum exchange between gas and dust
grains. The thermal radiation from the heated inner atmosphere and core
is sufficient to levitate dust grains and to push them outward.
Photophoresis can significantly accelerate the formation of giant
planets.}

\maketitle

\section{Introduction}

Up to 15\% of Sunlike stars may host a giant planet consisting of a
solid core and a gas envelope \citep{Howard2012}. Such planets require a
gaseous environment during their formation, as they must gravitationally
attract H$_2$ and He. Giant planet formation must therefore be completed
within the lifetime of the gaseous protoplanetary disks, 10~Myr at most
\citep{Meyer2007, Currie2009}.

The planet-metallicity correlation implies that bottom-up growth of
giant planets from solid cores, via core accretion, is the dominant
planet formation mechanism: planets prefer to orbit stars rich in iron,
a major grain-forming material \citep{Gonzalez1998, Fischer2005,
Neves2009}. Further studies of the composition of planet-host stars show
that the efficiency of dust grain nucleation is an important factor in
determining whether planets can form \citep{Robinson2006,
Brugamyer2011}. The first step in forming planets by core accretion is
the coagulation of small dust grains, which experiments have shown can
reach decimeter sizes simply due to van der Waals forces
\citep{Blum2008, Teiser2011b}. Dynamical concentration processes
such as the streaming instability then allow the dust aggregates to
reach large particle densities and form planetesimals
\citep{Johansen2007, Youdin2007}, at which point gravity becomes
the dominant force in planet growth. 

However, the small grains that are essential building blocks of the
giant planets become impediments to planet growth once the planet starts to accrete gas.
Small particles ($<$ mm) couple well to the surrounding gas, which
increases the opacity of the atmosphere severely \citep{Podolak2003}.
The accretion energy produced in the inner part of the growing planet is
absorbed by the dust grains, as the atmosphere gets optically thick for
thermal radiation. This heats up the system, which directly leads to an
increasing gas pressure. This pressure increase slows down the accretion
process, as the pressure gradient supports the gas against gravity.
Planet formation models have problems reproducing the fast timescales
required to explain giant planet formation \citep{Pollack1996}, and
those that succeed in forming planets on million-year timescales invoke
an arbitrary reduction of the opacity to 2\% of the interstellar value,
assuming that grains quickly settle to the sublimation point
\citep{Hubickyj2005, Lissauer2009, DodsonRobinson2010}.

There are multiple sources of dust in a protoplanetary
atmosphere. Accreted gas brings with it dust agglomerates
that are entrained in the gas \citep{Sicilia2011}. These primordial dust
particles should already be porous agglomerates, as coagulation of
single dust grains to small, highly porous agglomerates is a hit-and
stick mechanism which is very fast \citep{Dullemond2005, Blum2008}.
Porous particles have even higher absorbing cross-sections per unit mass
than the spheres on which the opacities used by \citep{Pollack1996} were
based, exacerbating the opacity problem. Planetesimals also drift into
the planetary core and are exposed to gas drag and to erosion during their
way through the planetary atmosphere. This erosion can already start in
the outer part of the atmosphere, as gas drag is an efficient erosion
mechansim even at low gas pressures \citep{Paraskov2006}. Intense
radiation from the inner part of the growing planet can add additional
erosion processes, as light induced thermodynamic processes can lead
to particle ejection from porous surfaces \citep{Kelling2011, DeBeule2013}. Erosion
of planetesimals generates large amounts of small dust aggregates, which
are entrained into the planetary atmosphere and can increase the
solid/gas ratio and therefore the opacity.

Within this work we present photophoresis as a mechanism capable of
pushing small particles outward and clearing the inner parts of the
atmosphere from dust grains. Photophoresis occurs when temperature
gradients caused by uneven illumination of a dust grain lead colliding
gas molecules to transfer momentum preferentially to the hotter, more
illuminated side of the dust grain. It has already been introduced as an
efficient mechanism for material transport in protoplanetary disks
\citep{Krauss2005, Wurm2006, Wurm2009}. \citet{Krauss2007} have also
shown that photophoresis is capable of clearing the inner zone of
protoplanetary disks of small grains. A growing giant planet is similar
to a protoplanetary disk in that in both cases an inner heat source
heats a gaseous environment by radiation, producing an outward flux
gradient. If photophoresis can eject dust particles from protoplanetary
atmospheres faster than they are replenished by accretion, it can
drastically reduce the atmospheric accretion timescale and increase the
gas/solid mass ratio of the resulting protoplanet
\citep{DodsonRobinson2008}. Here we demonstrate that the dust clearing
timescale for two snapshots of growing protoplanetary atmospheres is of
order $10^4$--$10^5$ years, far quicker than the planet growth
timescale.

In section \ref{model}, we discuss the model protoplanetary atmospheres
used for the photophoresis calculation. The principle of photophoresis
and the important parameters which determine its efficiency are
presented in section \ref{photo}. In section \ref{results} we
describe the resulting drift timescales from our photophoresis
model. In section \ref{discussion}, we discuss the efficiency of
photophoresis compared with other dust-clearing mechanisms and
present evidence that agglomerates in the protoplanetary
atmosphere should be porous, increasing the photophoretic force.
We present our conclusions in section \ref{conclusions}.

\section{Model}\label{model}

The growth of a protoplanet can be divided into three phases: (1)
planetesimal accretion that builds up a solid core, (2) slow gas and
planetesimal accretion that builds the protoplanetary atmosphere at a
nearly constant rate, and (3) runaway growth that takes the protoplanet
from ice-giant mass ($\sim 15 M_{\oplus}$) to gas-giant mass in $\sim
1000$~years \citep{Pollack1996, Hubickyj2005}. It is Phase 2 which
concerns us here---an epoch where cooling and contraction of the
protoplanetary atmosphere is limited by continual planetesimal heating
and high opacity due to small grains, preventing new gas from entering
the protoplanet's Hill sphere.

Here we present a proof-of-concept that photophoresis can affect the
thermal balance of a protoplanetary atmosphere by calculating the
photophoretic force in two Phase 2 time snapshots from the Uranus and
Neptune growth models of \citet{DodsonRobinson2010} for a protoplanet at
15~AU from the Sun. We use the planet-growth models as stand-alone input
into the photophoresis calculation, which was not incorporated into the
atmospheric growth model. 

In each time snapshot, the protoplanet atmosphere fills the smaller of
the Hill sphere, $ R_h = a[M_{planet} / (3 M_*)]^{1/3}$ (where $a$ is
the semimajor axis of the protoplanet's orbit, \citep{Armitage2010}), or the Bondi sphere,
$R_{acc} = G M_{planet} / c_s^2$ (where $c_s$ is the sound speed of the
surrounding nebula, \citep{Bondi1952}). The protoplanet is accreting 100-km planetesimals
at a rate of
\begin{equation}
\dot{M}_{solid} = C_1 \pi \Sigma_{solid} R_c R_h \Omega,
\label{dmdtcore}
\end{equation}
where $\Sigma_{solid}$ is the surface density of planetesimals in the
disk, $R_c \approx 0.01 R_h$ is the effective capture radius of the
protoplanet, which is well within the range found in literature \citep{Kary1993}. $\Omega$ is the Keplerian angular speed and $C_1$ is a
constant near unity \citep{Papaloizou1999}. Planetesimals enter the
protoplanet atmosphere at the Hill velocity, $v_h = R_h \Omega$, and
begin to deposit their kinetic energy in each layer of the atmosphere by
gas drag heating, dissipation of the kinetic energy of ablated material,
and sinking of material ablated in layers above. Energy is also removed
from the atmosphere due to latent heat as planetesimal material
vaporizes. For full details of the energy calculation, see
\citet{Pollack1996}.  A planetesimal is considered captured if it
displaces an amount of gaseous material equal to its own mass or
deposits 50\% or more of its mass into the atmosphere. Once 50\% of the
planetesimal mass has been ablated, we invoke the sinking approximation
so that the rest of the planetesimal and the ablated debris sink rapidly
to the core \citep{Hubickyj2005}.

%\begin{equation}
%\Delta E_i = F_d \Delta s_i + \frac{1}{2} \Delta m_i v_p^2 +
%\sum_{i' = 1}^i \Delta m_{i'} \left ( G M_i \left ( \frac{1}{R_i} -
%\frac{1}{R_{i=1}} \right ) \right ) - L.
%\label{kedeposit}
%\end{equation}
%In the first term, which where $F_d$ is the drag force on the planetesimal, $\Delta s_i$ is
%the is the path length through the $i$th layer of the
%atmosphere, $\Delta m_i$ is the total mass of the planetesimal
%vaporized or ablated in layer $i$, $v_p$ is the velocity of
%the planetesimal relative to the core, . Energy can also be
%deposited in layer $i$ when material ablated from layers above
%$i$ sinks through layer $i$.

We use a Henyey-type code to solve the standard equations of stellar
structure for the protoplanetary atmosphere \citep{Henyey1964}.
Opacities are modified from the standard interstellar case according to
the grain growth and settling prescription of \citet{Podolak2003}.  New
gas enters the atmosphere due to the combined effects of cooling-induced
contraction and expansion of the Hill sphere due to accretion.  During
each timestep of the planet-growth simulation, the atmosphere is assumed
to be in hydrostatic equilibrium. The first snapshot selected for the
photophoresis study (Snapshot 1) has atmospheric mass $M_{atm} = 0.11
M_{\oplus}$ and core mass $M_{core} = 4.39 M_{\oplus}$. Snapshot 2 has
$M_{atm} = 0.3 M_{\oplus}$ and $M_{core} = 7.48 M_{\oplus}$

\section{Photophoresis}\label{photo}

\begin{figure}[tb]
\center
\includegraphics[width=6cm]{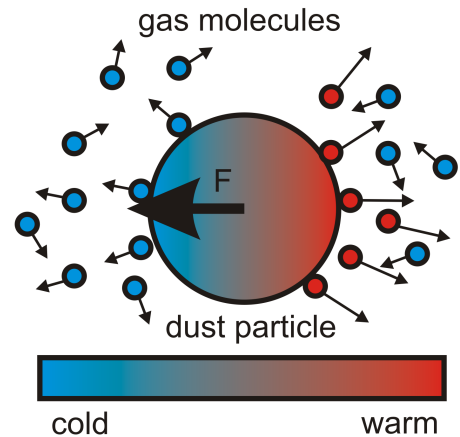}
\caption{Schematic sketch of the principle of photophoresis. The
illuminated surface heats up and transfers momentum to gas molecules,
which finally leads to a net force in the direction of light.}
\label{scheme}
\end{figure}

Photophoresis is an interaction between particles and the surrounding
gas in a medium with non-isotropic illumination. The illuminated side of
a particle is heated as the particle absorbs light. The other side stays
cooler and a temperature gradient forms along the particle surface. A
temperature gradient forms for any kind of radiation which is absorbed
by the particle (optical or thermal infrared). Gas molecules accommodate
to the particle surface and adopt the surface temperature where they
collide with a grain.  Such collisions change the mean velocity of the
molecules, as the mean thermal velocity depends on grain temperature as
$T_{grain}^{1/2}$. On the warm side of a grain, colliding gas molecules
gain more momentum than on the cooler side. The velocity difference
leads to a net momentum transfer from the gas molecules to the particle,
which amounts to a net force in the direction of the temperature
gradient (direction of light flux). The principle of photophoresis is
visualized in Figure \ref{scheme}.

Photophoresis is a thermodynamic non-equilibrium process and depends on
many particle and gas parameters, especially the gas pressure. It
reaches the maximum force $F_{max}$ at the pressure $p_{max}$, at which
the mean free path of the gas molecules is of the same order of
magnitude as the particle size. A semi-empirical description of the
photophoretic force $F_{ph}$ and the influence of the thermodynamic parameters and
particle size is given by \citet{Rohatschek1995}:
\begin{eqnarray}
F_{ph} & = & 2 F_{max} \left(\frac{p_{max}}{p} + \frac{p}{p_{max}}\right)^{-1}\label{pressure}\\
p_{max} & = & \frac{\eta}{r} \sqrt{\frac{12 R_{gas} T}{M_{mol}}}\label{pmax} \\
F_{max} & = & \frac{\pi \eta r^2 I}{2 k_{th}} \cdot \sqrt{\frac{R_{gas}}{3 T M_{mol}}}\label{fmax}\;.
\end{eqnarray}

Here, $r$ is the particle radius, $\eta = 1.8 \times 10^{-5} \; {\rm Pa \cdot s}$ is the dynamic gas viscosity, $k$ is the thermal
conductivity of the dust, $I$ is the incoming irradiation, $T$ is the
gas temperature, $M_{mol}$ is the molar mass of the gas (2.34 g/mol for
a mixture of mostly hydrogen and helium), and $R_{gas} = 8.31$ J/(mol K)
is the gas constant. \citet{Rohatschek1995} uses an additional thermal
creep parameter in the model which was set to 1.14. This is not
necessarily the case for real particles but to compare experimental
results with the semi-analytical solution by \citet{Rohatschek1995} it
is set to 1 for simplicity in this work.

The planetary atmospheres calculated in the models used here cover a
wide pressure range from several bar in the inner atmosphere to mbar in
the outer atmosphere. Although the pressures in the inner part are much
larger than the optimal pressure for photophoresis, the radiation flux
from the inner towards the outer parts is so large that photophoresis
has to be considered. To calculate the photophoretic force we assume
spherical dust agglomerates with a radius of $r = 10 \mathrm{\mu m}$,
thermal conductivity of $k_{th} = 0.1 \mathrm{W K^{-1} m^{-1}}$, and a
mean density of 2000 kg/m$^3$. This density corresponds to an olivine
agglomerates with a porosity of the order of 55 \%, which is still very
compact in comparison to dust agglomerates formed by coagulation
processes \citep{Blum2008}. For highly porous quartz agglomerates
thermal conductivities of $k_{th} \approx 0.01$ W/Km were found by
\citet{Krause2011} for similar porosities. We take a rather conservative
value of $k_{th}$ so as not to overestimate $F_{max}$. Our grain size of
$r = 10 \mathrm{\mu m}$, which is in the range of grain sizes determined
for comets \citep{Harker2002}, is also conservative as $p_{max}$
increases with decreasing particle size. All other parameters (light
flux, pressure, temperature) are taken from the formation model
described in section \ref{model}.

To evaluate the efficiency of photophoresis as a transport mechanism we
calculated the force ratio between photophoresis and gravity for
Snapshots 1 and 2. As long as the ratio between the two forces is larger
than 1, particles drift outward until they reach the point of
equilibrium. However, not only the force ratio is of great interest, but
also the the time the particles need to reach the point of equilibrium.
We assumed that solid particles exist as soon as the temperature is
below 1800 K and calculated the drift time they need to reach the point
of equilibrium.

Particles are accelerated by the acting resulting force $F_{res} =
F_{ph} - F_{grav}$ with $F_{grav} = \gamma \cdot m_{dust}\cdot m_{in}\cdot r^{-2}$ with the gravity constant $\gamma = 6.674^{-11} \mathrm{kg m^{-3}}$ and $m_{in}$ being the total mass of the core and the atmosphere within the radius $r$.  Particles are decelerated by gas drag, which leads to a
constant drift velocity of
\begin{equation}
v_{drift} = a_{res} \cdot \tau\,.
\end{equation} 
Here, $a_{res}$ is the resulting acceleration ($F_{res}/m_{dust}$) and $\tau$ is the gas
coupling time of the dust grains. The gas coupling behaviour of the dust
grains depends on the pressure. Hydrodynamical processes depend on the
Knudsen number, which is defined as $Kn =  \lambda / r_{dust}$ with the mean
free path $\lambda$ and the particle radius $r$. To determine the gas
coupling time for $Kn > 10$ we use the gas drag of free molecular flow
regime, which leads to
\begin{equation}
\tau = 0.68 \cdot \frac{m_{dust}}{\pi r_{dust}^2\rho_{gas}\nu_{gas}} \,=\, 0.91\cdot \frac{r_{dust} \rho_{dust}}{\rho_{gas}\nu_{gas}}\,.
\end{equation}
Here, $\rho_{gas}$ is the density and $\nu_{gas}$ is the thermal
velocity of the gas molecules, which can be written as $\nu_{gas} =
\sqrt{8R_{gas}T/(\pi M_{mol})}$ with the mass of a dust grain $m_{dust}$. The gas coupling time for $Kn < 0.1$ is
determined by the Stokes friction and is given as
\begin{equation}
\tau = \frac{m_{dust}}{6\pi r_{dust} \eta} \, = \, \frac{2 r_{dust}^2 \rho_{dust}}{9 \eta}\,.
\end{equation}
Stokes friction can be applied since the Reynolds numbers for dust
particles are well below $Re = 1$, even for the large densities in the
inner atmosphere. For the transition regime with $0.1 \leq Kn \leq 10$
we use Stokes friction with the Cunningham correction
\citep{Cunningham1910, Hutchins1995}, which is given as
\begin{equation}
\tau = \frac{m_{dust}}{6\pi r_{dust} \eta} \cdot \left(1 + Kn \cdot \left(1.231 + 0.47 e^{-1.178 / Kn}\right)\right).
\end{equation} 
With this set of equations the coupling times and therefore also the
drift velocities can be calculated for dust grains moving outward, which
also gives the drifting timescales for the dust grains.

\section{Results}\label{results}

\begin{figure}[tb]
\includegraphics[width=9cm]{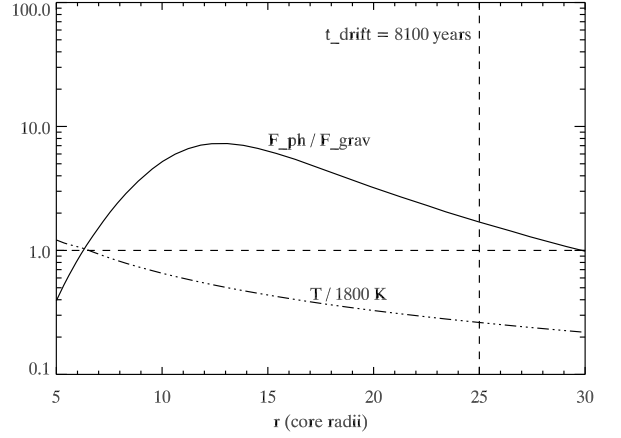}
\caption{Photophoretic strength and temperature distribution for a
growing giant planet with a total mass (core+atmosphere) of 4.5 earth
masses and a core radius of 1.64 earth radii.}
\label{4.5Me}
\end{figure}

The accretion rate and therefore also the heat production in the inner
part of the growing atmosphere change during the formation of giant
planets. Within this work the influence of photophoresis on the
evolution of giant planets is studied for a protoplanet at 15 AU solar
distance at two different evolutionary stages. Snapshot 1 presents the
growing planet after a formation time of 3.07 million years with a
planetary core of 4.39 earth masses and an atmosphere of 0.11 earth
masses. The result is shown in Figure \ref{4.5Me}.

Dust particles only exist in the regions of the planetary atmosphere, in
which the temperature (dash-dotted curve, normalized to 1800 K) is below
the condensation temperature (we assumed 1800, although in reality dust
grains condense over a range of approximately 1200 K-2100 K). In
Snapshot 1, dust grains begin to condense at a radial distance from the
center of the protoplanet of 6.4 core radii. At this point the
photophoretic force (solid curve, normalized by gravity) is larger than
gravity by a factor of 1.1. The point of equilibrium, for which the
photophoretic force exactly balances gravity, is at a radial
distance of 30 core radii. A dust grain with a radius of $r =
10\,\mathrm{\mu m}$ needs $3 \cdot 10^4$ years to drift the distance
between those points. It is important to note that the drift process
takes longest in the outer part of the atmosphere, as it takes only 8100
years for the same dust grain to reach a radial distance of 25 core
radii (see Figure \ref{4.5Me}).

\begin{figure}[tb]
\includegraphics[width=9cm]{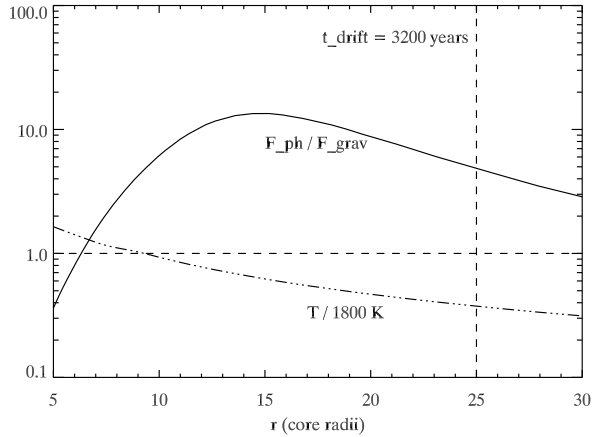}
\caption{Photophoretic strength and temperature distribution for a growing giant planet with a core of 7.8 earth masses and a core radius of 1.96 earth radii.}
\label{7.8Me}
\end{figure}

In Snapshot 2, a later growth stage, the growing planet has accreted
much more mass and more accretion heat has developed in the inner part,
so temperatures increase and photophoresis gets stronger. The result is
shown in Figure \ref{7.8Me} for a planetary core of 7.48 earth masses
and an atmosphere of 0.3 earth masses with a formation time of 4.39
million years.  The region in which dust grains exist has moved outward
to a radial distance of 11.1 core radii, while the point of equilibrium
between photophoresis and gravity has moved to 42 core radii. At the
inner edge of the dusty part of the atmosphere photophoresis exceeds
gravity by a factor of 4, which leads to drift timescales of $3.3
\cdot 10^4$ years from this point to the point of equilibrium. Here, it
takes only $3.2 \cdot 10^3$ years for the same dust agglomerate to reach
a radial distance of 25 core radii. In both snapshots photophoresis is
capable of moving dust outward.

\section{Discussion}\label{discussion}

 An additional radiation-based process is radiation pressure,
which is also capable of accelerating particles. Radiation pressure is
the momentum transfer from photons on a particle and can be written as
$F_{rad} = r^2 \cdot I/c$ for a spherical, absorbing particle, with the
intensity $I$, light speed $c$ and the particle radius $r$. Radiation
pressure plays a significant role in the solar system, as comet tails
and asteroids are strongly influenced by it or by related processes
(Poynting-Robertson Effect, Yarkowsky Effect). For gaseous environments
like the solar nebula or protoplanetary atmospheres, the photophoretic
force exceeds radiation pressure by orders of magnitude. For both snapshots (Figures
\ref{4.5Me} and \ref{7.8Me}) photophoresis is larger than radiation
pressure by at least three orders of magnitude for the dust-rich part of
the planetary atmosphere.

In equation \ref{fmax} it is shown that photophoresis scales directly with the intensity of the light flux. This parameter varies strongly with the radial distance to the planetary core and depending on the planet formation scenario. In case of the second snapshot (Figure \ref{7.8Me}) an intensity reduced by a factor of 10 will reeduce the zone in which photophoresis exceeds gravity to a radial distance from the core between 12 and 19 core radii. In case of an intensity by a factor of 10 larger than in Figure \ref{7.8Me}, this zone extends from the direct core environment to a distance of 90 core radii instead of 42 core radii as in Figure \ref{7.8Me}. 

The presented results show that photophoresis can indeed be an
important process for the formation of giant planets. Depending
on the evolutionary state of the growing planet, photophoresis
can exceed gravity by a factor of 10, leading to an outward
drift of dust grains. The timescales for this drift process are
small in comparison to the formation timescales of giant
planets. The gas pressure within the inner part of the planetary
atmosphere is much larger than the pressure for which
photophoresis is most efficient. The photophoretic efficiency
will therefore increase with the growing distance to the
planetary core of the grain-sublimation line. 

The light intensity drops steeply with the radial distance, which leads
to decreasing photophoresis, especially as dust in the planetary
atmosphere absorbs and scatters the radiation from the innermost
atmosphere, enhancing the decrease. However, as dust grains drift
outward due to photophoresis, the zone where $F_{ph} > F_{grav}$ is
quickly cleared from absorbing or scattering dust grains. This clearing
leads to an even stronger effect of photophoresis, as the radiation from
the inner parts can then reach the dust grains directly. A similar
mechanism has already been described for the inner zone of
protoplanetary disks by \citet{Krauss2007}. The consequence of this
mechanism is a kind of runaway process, as the clearing of the inner
zone enhances photophoresis in the outer parts of the planetary
atmosphere. This feedback reaction is not treated in the snapshots shown
in Figures \ref{4.5Me} and \ref{7.8Me}, as a revision of the complete
formation model is beyond the scope of this work. In future work we will include photophoresis
in the atmospheric opacity calculations, so the atmospheric accretion
rate and photophoretic force can be calculated self-consistently.

The material properties of the dust agglomerates in the planetary
atmosphere are based on simplified assumptions concerning density,
thermal conductivity, and porosity. It is important to note that these
properties have a strong influence on the photophoretic acceleration.
The exact properties of the dust grains are unknown. Bodies produced by
coagulation processes are highly porous ($\geq 60 \%$ porosity), even
if the collision velocities reach values of 60 m/s \citep{Teiser2009a,
Meisner2012}. Planetesimals, which are not thermally processed, are
still highly porous. Fragments which are generated by erosion processes
will therefore also be highly porous. Primordial dust generated by
condensation in protoplanetary disks is initially micron-sized during
the early formation phase of the disks. Micron-sized particles coagulate
very efficiently, as collision velocities are small and cohesion forces
are large. Efficient coagulation has been demonstrated both
theoretically and experimentally \citep{Krause2004, Wada2009}. Single
monomers do therefore not exist after very short timescales due to this
fast coagulation \citep{Dullemond2005}. In a more complete calculation
of the photophoretic force, dust particles should be treated as highly
porous agglomerates rather than spherical, porous monomers. 

The thermal properties of porous agglomerates are strongly influenced by the porosity. With a
porosity of 55 \% and $k_{th} = 0.1 \mathrm{W/Km}$ we considered rather
small porosities and large thermal conductivities in comparison to
literature \citep{Meisner2012, Krause2011}. The photophoretic force
scales with $1/k_{th}$, so a greater porosity leads directly to a
smaller thermal conductivity and therefore to an increased photohoretic
force. The reduced density automatically leads to a larger acceleration
due to smaller inertia. In this case the clearing process will be
enhanced dramatically. Of course, for perfectly solid, non-porous
monomers the clearing efficiency will be reduced and photophoresis will
not exceed gravity in the inner part of the atmosphere. However, it is
questionable if such solid particles exist in large amounts in the outer
part of the solar system. Solid particles of mm-size, like e.g.
chondrules, usually are thermally processed \citep{Scott2007}, which is
assumed to happen in the inner part of the solar system. Bodies in the
outer solar system typically contain large amounts of ice and are
assumed to be highly porous. We conclude that the majority of the small
particles in the atmosphere of growing giant planets consists of porous
dust agglomerates with very small thermal conductivities.

Mutual collisions between dust agglomerates might lead to a fragmentation cascade which leads to a wide size distribution. In both snapshots photophoresis is capable of transporting material of a certain size distribution, but the timescales vary significantly with the particle size. In the first snapshot (Figure \ref{4.5Me}) the drift timescale to reach the point of equilibrium increases to $2.6 \cdot 10^5$ for a particle radius of $1 \mu\mathrm{m}$. Additionally, the starting point, for which photophoresis exceeds gravity also varies, as the photophoretic force reaches its maximum at different pressures depending on the particle size. With increasing particle size this thresholds is found at larger distances from the planetary core, which means that for 100 $\mu$m particles this mechanism could only work at a radial distance between 23 core radii and 25 core radii. In the second snapshot (Figure \ref{7.8Me}) photophoresis exceeds gravity in the complete inner part of the dust rich atmosphere for particle sizes up to 200 $\mu$m. Howver, the timescales still vary from $t = 3 \cdot 10^5$ years for a particle radius of $1 \mu$m to $t = 7 \cdot 10^3$ years for a particle radius of $100 \mu$m.    

Photophoresis is based on the non-symmetric heating of particles, as a temperature gradient along the surface has to form. In case of fast rotation particles are heated more homogeneously, which reduces the temperature gradient. In case of planetary atmospheres the gas pressure is high, which leads to a strong damping of any rotation. Additionally, experiments have shown that particles tend to align their rotation axis parallel to the light flux, if they are subject to photophoresis and gravity \citep{vaneymeren2012}. A temperature gradient in the direction of the light flux is therefore not prevented. Both processes lead to the conclusion that rotation does not prevent photophoretic motion and can be neglected in simplified calculations as this. 

\section{Summary and Conclusions}\label{conclusions}

The formation of giant planets has to happen fast, as their formation
has to be completed within the lifetime of the protoplanetary disk.
Small particles in the planetary atmosphere reduce the formation speed,
as accretion-produced heat cannot be efficiently transported outward by
radiation.  Photophoresis can push small particles to the outer parts of
the planetary atmosphere, clearing the inner parts so accretion heat can
be quickly transported outward. In the cleared zone no absorption takes
place, so the light flux in the outer regions increases as well.
Photophoretic clearing therefore leads directly to a runaway process, in
which photophoresis amplifies itself by clearing the inner parts of the
atmosphere.

As the heat is transported outward, the forming protoplanet begins to
cool and contract efficiency, allowing new gas to enter the Hill sphere.
Photophoresis therefore increases the possible gas accretion rate and
may lead to a substantial speedup in giant planet formation.
\citet{DodsonRobinson2008} demonstrated that a low-opacity atmosphere increases the gas/solid mass ratio of the newly formed planet, so
photophoresis may modify planetary structure as well as growth
timescale. In future work we will present updated planet-growth
calculations that self-consistently incorporate photophoresis into the
planetary atmosphere calculation.

\begin{acknowledgements}
We thank Peter Bodenheimer for the collaboration and for providing access to the planet formation model. 
Funding for S.D.R.'s work was provided by National Science
Foundation CAREER award AST-1055910.
\end{acknowledgements}

%\include{giantplanets_bib}
%\begin{thebibliography}{}
%\bibliographystyle{aa} % style aa.bst
%\bibliography{references.bib} % your references Yourfile.bib

%\end{thebibliography}

\clearpage

\end{document}